\pdfoutput=1

\documentclass[aps,prd,amsmath,superscriptaddress,%
               nobibnotes,nofootinbib,preprintnumbers,
               onecolumn,12pt,tightenlines]{revtex4}

\usepackage{ifthen}

\usepackage{dcolumn}

\newcommand{\refeqn}[2][eqn:]{Eqn.~(\ref{#1#2})}

\newcommand{\reftab}[2][tab:]{Table~\ref{#1#2}}
\newcommand{\Reftab}[2][tab:]{Table~\ref{#1#2}}

\newcommand{\ifmulticol}[2]{%
  \ifthenelse{\lengthtest{1.9\columnwidth<\textwidth}}{#1}{#2}%
}

\newcommand{\orderof}[1]{\ensuremath{\mathcal{O}(#1)}}

\newcommand{\gae}{%
  \ensuremath{\lower 2pt \hbox{%
    $\, \buildrel {\scriptstyle >}\over {\scriptstyle \sim}\,$}%
    }%
  }
\newcommand{\lae}{%
  \ensuremath{\lower 2pt \hbox{%
    $\, \buildrel {\scriptstyle <}\over {\scriptstyle \sim}\,$}%
    }%
  }

\newcommand{\mchi}{\ensuremath{m_{\chi}}}
\newcommand{\rhochi}{\ensuremath{\rho_{\chi}}}
\newcommand{\nchi}{\ensuremath{n_{\chi}}}

\newcommand{\mbody}{\ensuremath{M_{\mathrm{body}}}}

\newcommand{\qmax}{\ensuremath{q_{\mathrm{max}}}}

\newcommand{\fpSI}{\ensuremath{f_{\mathrm{p}}}}
\newcommand{\fnSI}{\ensuremath{f_{\mathrm{n}}}}
\newcommand{\apSD}{\ensuremath{a_{\mathrm{p}}}}
\newcommand{\anSD}{\ensuremath{a_{\mathrm{n}}}}

\newcommand{\sigmaSI}{\ensuremath{\sigma_{0,\mathrm{SI}}}}
\newcommand{\sigmaSD}{\ensuremath{\sigma_{0,\mathrm{SD}}}}

\newcommand{\sigmapSI}{\ensuremath{\sigma_{\mathrm{p,SI}}}}

\newcommand{\sigmapSD}{\ensuremath{\sigma_{\mathrm{p,SD}}}}

\newcommand{\mup}{\ensuremath{\mu_{\mathrm{p}}}}

\newcommand{\Sp}{\ensuremath{\langle S_{\mathrm{p}} \rangle}}
\newcommand{\Sn}{\ensuremath{\langle S_{\mathrm{n}} \rangle}}

\begin{document}




\title{Dark Matter collisions with the Human Body}

\author{Katherine Freese}
\email[]{ktfreese@umich.edu}
\affiliation{
 Michigan Center for Theoretical Physics,
 Department of Physics,
 University of Michigan,
 Ann Arbor, MI 48109}

\author{Christopher Savage}
\email[]{savage@fysik.su.se}
\affiliation{
 The Oskar Klein Centre for Cosmoparticle Physics,
 Department of Physics,
 Stockholm University,
 AlbaNova,
 SE-106 91 Stockholm, Sweden}

\date{\today}



\begin{abstract} 


We investigate the interactions of Weakly Interacting Massive Particles
(WIMPs) with nuclei in the human body.  We are motivated by the fact that
WIMPs are excellent candidates for the dark matter in the Universe.
Our estimates use a 70~kg human and a variety of WIMP masses and
cross-sections.  The contributions from individual elements in the body
are presented and it is found that the dominant contribution is from
scattering off of oxygen (hydrogen) nuclei for the spin-independent
(spin-dependent) interactions.  For the case of 60~GeV WIMPs, we find
that, of the billions of WIMPs passing through a human body per
second, roughly $\sim 10$ WIMPs hit one of the nuclei in the human body
in an average year, if the scattering is at the maximum consistent with
current bounds on WIMP interactions. 
We also study the 10--20 GeV WIMPs with much larger cross-sections
that best fit the DAMA, COGENT, and CRESST data sets and find much
higher rates: in this case as many as $10^5$ WIMPs hit a nucleus in the
human body in an average year, corresponding to almost one a minute.
Though WIMP interactions are a source of radiation in the
body, the annual exposure  is negligible compared to that from other
natural sources (including radon and cosmic rays), and the WIMP
collisions are harmless to humans.

\end{abstract} 

\maketitle



A variety of astrophysical observations has shown conclusively that the
majority of the matter in the Universe consists of an unknown
nonluminous, nonbaryonic component. Understanding the nature of this
dark matter is one of the major outstanding problems of astrophysics
and particle physics.  Most cosmologists believe that the solution to
this puzzle lies in the discovery of a new type of fundamental particle.
Leading candidates for the dark matter are Weakly Interacting Massive
Particles (WIMPs), a generic class of particles that are electrically
neutral and do not participate in strong interactions, yet have
weak interactions with ordinary matter.  Possible WIMP candidates
include supersymmetric particles and Kaluza-Klein particles motivated
by theories with extra dimensions.  These particles are thought to have
masses in the range 1~GeV--10~TeV, consistent with their being part
of an electroweak theory.

Searches for WIMPs \cite{Goodman:1984dc,Drukier:1986tm,Freese:1987wu}
include direct detection laboratory experiments,
which look for the elastic scattering of WIMPs in the Galaxy as they
pass through terrestrial detectors situated in deep underground sites.
These efforts are ongoing worldwide.  Currently there are intriguing
hints of discovery with the DAMA \cite{Bernabei:2010mq},
CoGeNT \cite{Aalseth:2010vx,Aalseth:2011wp}, and
CRESST \cite{Angloher:2011uu} experiments although no consensus has
been reached in the community.  The null results of a host of other
experiments, including CDMS \cite{CDMSII} and
XENON \cite{Angle:2009xb,Aprile:2011hi} have been used to place
bounds on the scattering rates of WIMPs as a function of WIMP mass.
In the standard framework used in this work, there is a strong tension
between the results of the first three experiments and the null results
of the latter two.
Many efforts in both the experimental and theoretical directions are
ongoing to understand these discrepancies;  in this paper we will simply
use the currently published results of these experiments.

In this paper we consider this same elastic scattering of WIMPs with
nuclei in the human body.  Billions of WIMPs pass through our bodies
every second, yet most of them pass through unimpeded. Only rarely does
WIMP actually hit one of our nuclei.  To perform our analysis
we will assume a human of 70~kg and consider a variety of WIMP masses
in the GeV--TeV range.
First we will study 60~GeV WIMPs with the maximum scattering
cross-section allowed by the null results of the XENON and CDMS
experiments.
Then we will turn to the lower mass WIMPs (10--20 GeV) that provide the
best fits to the hints of discovery in DAMA, CRESST, and COGENT as well
as TeV benchmark cases again compatible with the null result
experiments.
Finally, we examine the radiation exposure these interactions represent
and how it compares to other natural radiation sources.


The scattering rate of WIMPs with an element (indexed by $k$) in a
human body of mass $\mbody$ is given by\footnote{
    The rate here is a pure rate, not a rate per unit target mass as is
    commonly used in the dark matter direct detection literature.}
\begin{equation}\label{eqn:dRdE}
  R_k = N_k \nchi \langle v  \sigma_k \rangle
      = \left( \frac{f_k \mbody}{m_k} \right)
        \left( \frac{\rhochi}{\mchi} \right)
        \int d^3v \, v f(\mathbf{v}) \sigma_k(v) \, ,
\end{equation}
where $N_k = \frac{f_k \mbody}{m_k}$ is the number of nuclei of that
element in the body, with $m_k$ the nuclear mass and $f_k$ the mass
fraction of that element;
$\nchi = \frac{\rhochi}{\mchi}$ is the number density of WIMPs, with
$\mchi$ the WIMP mass and $\rhochi$ the local dark matter mass density;
$f(\mathbf{v})$ is the WIMP velocity distribution; and
$\sigma_k(v)$ is the (velocity-dependent) WIMP-nucleus scattering
cross-section.

To a reasonable first approximation, the dark matter halo can be treated
as a non-rotating, isothermal sphere (the Standard Halo Model)
\cite{Drukier:1986tm,Freese:1987wu}.
For the resulting Maxwellian velocity distribution, a 3D velocity
dispersion of 270~km/s is assumed.  The velocity distribution is
truncated at 550~km/s to account for the fact that high velocity
particles would escape the galaxy \cite{Smith:2006ym}, though the
results of this paper are fairly insensitive to this cutoff as such
high velocity particles would otherwise make only a small contribution
to the \textit{total} scattering rate\footnote{%
    The same cannot always be said for the rates in direct detection
    experiments as these experiments are sensitive to events that
    produce energies above a threshold, not the total number of events.
    In some cases, only high velocity WIMPs produce scattering events
    above threshold, so the choice of cutoff becomes important.
    }.
The local density of the dark matter halo is taken to be 0.4~GeV/cm$^3$.
While the smooth halo component is likely to be supplemented by a
variety of substructures such as streams, clumps, or debris flow, their
contributions are unlikely to be large enough to substantially modify
the results of this paper.

Dropping the isotope index $k$, the scattering cross-section is given by
\begin{equation}\label{eqn:sigma}
  \sigma(v) = \int_0^{\qmax^2} dq^2 \frac{d\sigma}{dq^2}(q^2,v) \, ,
\end{equation}
where $q$ is the momentum transferred in a scatter, $\qmax = 2 \mu v$ is
the maximum momentum transfer in a scatter at a relative velocity $v$,
$\mu$ is the WIMP-nucleus reduced mass, and
\begin{equation}\label{eqn:dsigma}
  \frac{d\sigma}{dq^2}(q^2,v) = \frac{\sigma_{0}}{4 \mu^2 v^2}
                              F^2(q) \, \Theta(\qmax-q) \, ,
\end{equation}
with $\Theta$ the step function and $\sigma_0$ the scattering
cross-section in the zero-momentum-transfer limit.
Here, $F^2(q)$ is a form factor to account for the finite size of the
nucleus.  For small momentum transfers, the WIMP coherently scatters off
the entire nucleus; the nucleus is essentially a point particle in this
case, with $F^2(q) \to 1$.  For sufficiently small $v$, such that the
possible momentum transfer remains small, $\sigma(v) \to \sigma_0$.
As the de Broglie wavelength of the momentum transfer becomes
comparable to the size of the nucleus, the interaction becomes
sensitive to the spatial structure of the nucleus and $F^2(q) < 1$,
with $F^2(q) \ll 1$ at higher momentum transfers.  For velocities at
which this form factor becomes relevant, $\sigma(v) < \sigma_0$ (with
$\sigma(v) \ll \sigma_0$ at very high velocities).  The velocity at
which this form factor causes the cross-section $\sigma(v)$ to start to
significantly deviate from the zero-momentum-transfer limit $\sigma_0$
is dependent on the nuclei in question for two reasons: (1) the size of
the nucleus grows as the nucleus gets heavier and (2) the momentum
transferred becomes larger as the nucleus gets heavier, assuming the
WIMP is heavier than the nuclei in question.
For the typical WIMP velocities in the halo, the form factor suppression
is negligible for nuclei much lighter than iron ($\sigma(v) \approx
\sigma_0$), while it is significant for nuclei much heavier\footnote{%
    For the same reasons as given in the previous footnote (the
    application of a threshold), the form factor is more important for
    direct detection and it can significantly suppress the direct
    detection rates above threshold even when the total rate does is
    not significantly affected.
    }.

There are two types of interactions commonly considered for WIMP
scattering: spin-independent (SI) and spin-dependent (SD).  Each
coupling has its own form factor and \refeqn{dRdE} must be summed
over these two contributions.  In the SI case, the WIMP essentially
couples to the mass in the nucleus, with a zero-momentum-transfer limit
cross-section
\begin{equation} \label{eqn:sigmaSI}
  \sigmaSI = \frac{4 \mu^2}{\pi} \left[ Z \fpSI + (A-Z) \fnSI \right]^2 \, ,
\end{equation}
where $\fpSI$ and $\fnSI$ are the couplings to the proton and neutron,
respectively, $Z$ is the number of protons in the nucleus, and $A-Z$
is the number of neutrons.  For many WIMP candidates,
$\fpSI \approx \fnSI$ and the cross-section scales as
\begin{equation} \label{eqn:sigmaSI2}
  \sigmaSI = \frac{\mu^2}{\mup^2} A^2 \, \sigmapSI \, ,
\end{equation}
where $\mup$ is the WIMP-proton reduced mass and $\sigmapSI$ is the
SI WIMP-proton scattering cross-section.  We will assume $\fpSI = \fnSI$
below, though the results are only very mildly sensitive to the ratio
of these two couplings except in the case $\frac{\fpSI}{\fnSI} \approx
-\frac{A-Z}{Z}$ where the terms in \refeqn{sigmaSI} cancel.

In the SD case, as the name implies, the WIMP couples to the spin of
the nucleus, with
\begin{equation} \label{eqn:sigmaSD}
  \sigmaSD = \frac{32 \mu^2}{\pi} G_{F}^{2} J(J+1) \Lambda^2 \, ,
\end{equation}
where $J$ is the spin of the nucleus,
\begin{equation} \label{eqn:Lambda}
  \Lambda \equiv \frac{1}{J} \left( \apSD \Sp + \anSD \Sn \right) \, ,
\end{equation}
$\apSD$ and $\anSD$ are the couplings to the proton and neutron,
respectively, and $\Sp$ and $\Sn$ are the spin contributions
from the proton and neutron groups, respectively.
In our analysis, we shall assume identical couplings to the proton
and neutron ($\apSD = \anSD$), so that
\begin{equation} \label{eqn:sigmaSD2}
  \sigmaSD = \frac{\mu^2}{\mup^2}
             \frac{J(J+1)}{\frac{1}{2}(\frac{1}{2}+1)}
             \left( \frac{\Sp+\Sn}{J} \right)^2
             \, \sigmapSD \, .
\end{equation}
Whereas the couplings to neutrons and protons are roughly identical for
SI scattering for many WIMP candidates, in the case of SD scattering
they may differ.  Typically, however, the two SD couplings are found to
be within a factor of 2--3 of each other.
Our results, using identical couplings, will thus be order of magnitude
estimates of the general case.

More detailed discussions of dark matter scattering kinematics,
cross-sections, and form factors can be found in
Refs.~\cite{Lewin:1995rx,Jungman:1995df,Bednyakov:2004xq,Bednyakov:2006ux};
other reviews can be found in Refs.~\cite{Primack:1988zm,Bertone:2004pz}.


\begin{table}
  \begin{center}
  \newcolumntype{e}{D{.}{.}{5}}
  \newcolumntype{f}{D{.}{.}{5}}
  \addtolength{\tabcolsep}{0.75em}
  \begin{tabular}{leef}
    \hline \hline 
    \textbf{Element}
      & \multicolumn{1}{c}{\textbf{Mass}}
      & \multicolumn{2}{c}{\textbf{Rates [yr$^{-1}$]}} \\
    \textbf{}
      & \multicolumn{1}{c}{\textbf{Fraction}} 
      & \multicolumn{1}{c}{\textbf{SI}}
      & \multicolumn{1}{c}{\textbf{SD}} \\
    \hline 
    Oxygen     & 0.61    & 3.49    &  0.25   \\
    Carbon     & 0.23    & 0.63    &  0.64   \\
    Hydrogen   & 0.10    & 0.00023 & 22.5    \\
    Nitrogen   & 0.026   & 0.11    &  0.0097\makebox[0pt][l]{$^\dag$} \\
    Calcium    & 0.014   & 0.64    &  0.011  \\
    Phosphorus & 0.011   & 0.30    &  5.7    \\
    Potassium  & 0.0020  & 0.089   &  0.27   \\
    Sulfur     & 0.0020  & 0.059   &  0.0027 \\
    Sodium     & 0.0014  & 0.019   &  0.58   \\
    Chlorine   & 0.0012  & 0.043   &  0.079  \\
    Magnesium  & 0.00027 & 0.0043  &  0.024  \\
    Silicon    & 0.00026 & 0.0057  &  0.0023 \\
    Iron       & 0.00006 & 0.0050  &  0.00001\\
    \hline 
    \textit{Total}
               & 1.00    & 5.39    & 30.1    \\
    \hline \hline 
  \end{tabular}
  \end{center}
  \caption[Composition of the human body]{%
    Interactions of 60~GeV WIMPs on various nuclei in the human body.
    The mass fraction of the most significant elements in the human
    body, taken from Ref.~\cite{CRC:2012} (which in turn refers to
    Refs.~\cite{Padikal:1981,Snyde:1975}), is shown.
    Also shown are the number of WIMP scatters per year for each
    element at the largest spin-independent (SI) and
    spin-dependent (SD) scattering cross-sections not currently
    excluded by XENON100 \cite{Aprile:2011hi}, which are
    $\sigmapSI = 10^{-8}$~pb and $\sigmapSD = 2 \times 10^{-3}$~pb,
    respectively.
    We assume a human mass of 70~kg and identical couplings to the
    proton and neutron.
    ($\dag$) The SD rate for nitrogen-14 has not been calculated but may
    be non-negligible and perhaps as large as $\orderof{10}$;
    see the text.
    }
  \label{tab:Elements}
\end{table}

\Reftab{Elements} shows the mass fractions of the most significant
elements in the human body as well as the scattering rates for each
element for a 70~kg body and a 60~GeV WIMP.  Rates are shown for both
SI and SD scattering, assuming scattering cross-sections of 
$\sigmapSI = 10^{-8}$~pb and $\sigmapSD = 2 \times 10^{-3}$~pb,
respectively, the largest cross-sections not excluded by XENON at that
WIMP mass.
Oxygen and carbon are the largest components in the human body by mass
and also contribute the most to the SI scattering rate, with oxygen
accounting for 65\% of the SI scatters at this WIMP mass.  However,
hydrogen, the largest component by number of atoms (representing about
60\% of the atoms in the human body), has a much smaller SI scattering
rate than many other elements with significantly smaller mass fractions
(as well as number of atoms).  For example, iron, while accounting for
less than 1/1000 the mass of the hydrogen, nevertheless has an SI
scattering rate $\sim$20 times larger.  The reason for this lies in the
scaling of the SI cross-section shown in \refeqn{sigmaSI2}.  In addition
to the explicit $A^2$ factor, the $\frac{\mu^2}{\mup^2}$ factor also
scales as $A^2$ (for nuclei much lighter than the WIMP), so that the
cross-section scales as $A^4$.  For a given mass fraction, the number
of nuclei is proportional to $1/A$, so the interaction rate scales as
$A^3$.  With this scaling and the mass fractions shown in the table,
the relative oxygen-to-hydrogen SI scattering rate should then
approximately be $\frac{0.61}{0.10} \left(\frac{16}{1} \right)^3
\approx 25,000$, in reasonable agreement with the actual value of
$\frac{3.49}{0.00023} \approx 15,000$; the overestimate in the first
case is due to the fact that $\frac{\mu^2}{\mup^2} \to A^2$ applies in
the limit that the WIMP is much heavier than the nucleus, a limit that
has not been fully reached here.
As the nuclei become heavier, the form factor becomes more and more
significant, so the $A^3$ scaling in the interaction rate for a given
mass fraction no longer holds, though the rate still grows rapidly.

On the other hand, scattering with hydrogen is the dominant
contribution in the SD case.  The primary difference is that, unlike
the SI case, there is no explicit $A^2$ scaling in the scattering
cross-section: the spin factors in \refeqn{sigmaSD2} are of
$\orderof{1}$ for all nuclei.  With the $\mu^2$ factor, the SD
cross-section scales as $\sim A^2$.  After accounting for the $1/A$
scaling of the number of nuclei for a given mass fraction, the total
scattering rate scales as $\sim A$ (neglecting form factors).  However,
isotopes with zero nuclear spin ($J=0$) have $\sigmaSD = 0$, so they do
not contribute at all to the SD scattering rate.  Many of the elements
listed in \reftab{Elements}, including oxygen and carbon, are mainly
composed of spinless isotopes, with non-zero spin isotopes representing
only a small fraction of that element's natural composition.  The SD
scattering rate is thus suppressed in these cases.  Hydrogen, on the
other hand, is mainly composed of spin-1/2 $^1$H; even spin-1 deuterium
contributes to SD scattering.  Because of the $A$ scaling of the
scattering rate for a given mass fraction and the relative isotopic
compositions between spinless and non-zero spin nuclei, hydrogen
dominates the SD capture rate.

In our analysis, we have neglected the SD contribution of
spin-1 $^{14}$N.  As this is the dominant isotope of nitrogen, nitrogen
is expected to have a significant SD scattering rate.
However, this nucleus belongs to a small group of proton-odd, neutron-odd
isotopes with non-zero spin that are not well characterized in the
scattering literature and we are unaware of existing estimates
for $\Sp$ and $\Sn$.  Taking $|\Sp| \sim |\Sn| \sim 0.1$, similar to
nearby nuclei (except one of these two quantities is nearly zero in
these other nuclei), we can expect $\orderof{10}$ SD scattering events
per year with nitrogen in the human body.  This would make nitrogen one
of the larger contributors to the total SD rate, though hydrogen still
remains the dominant source of SD interactions.

The overall scattering rates of $\orderof{10}$ should not be unexpected
for the benchmark WIMP mass and cross-sections here.  These benchmarks
would produce a few events/year in the $\sim$100~kg of liquid xenon
that is the target mass in the XENON experiment, the currently measured
event rate in the detector (though the measured rate is also consistent
with backgrounds alone).
With a similar mass between the human body
and the XENON detector, the rates should be of similar orders of
magnitude, though detection efficiencies, thresholds, and different
target elements mean the rates are not simply proportional to the
target mass.
Since xenon ($A \approx 130$) is much heavier than
oxygen ($A \approx 16$), one might expect a much higher rate in XENON
than the human body for SI scattering due to the $\sim A^4$ cross-section
scaling ($\sigma_{0,\mathrm{Xe}}$ is $\orderof{10^3}$ larger than
$\sigma_{0,\mathrm{O}}$).  However, due to a threshold and a $<\!\!100\%$
detection efficiency, the few events/year rate measured in XENON is
not the \textit{total} rate in the detector, which is somewhat higher
(by an order of magnitude or more).  In addition, xenon scattering will
be form factor suppressed, so that the total scattering rate for xenon
is not as high as would be expected from the $A^4$ scaling alone.
For the SD case, the $\orderof{10}$ higher scattering rate in the human
body versus the XENON experiment can be attributed to the much larger
number of non-zero spin nuclei in the former case (mainly hydrogen).

\begin{table}
  \begin{center}
  \addtolength{\tabcolsep}{0.75em}
  \begin{tabular}{lccc}
    \hline \hline 
    \textbf{Benchmark}
      & \textbf{WIMP Mass}
      & \textbf{Cross-section}
      & \textbf{Rate} \\
    \textbf{}
      & \textbf{[GeV]}
      & \textbf{[pb]}
      & \textbf{[yr$^{-1}$]} \\
    \hline 
    \multicolumn{4}{c}{\textit{spin-independent}} \\
    CoGeNT best-fit
        &    8.  & 7.  $\times 10^{-5}$ & 6.3 $\times 10^{4}$ \\
    CRESST M1
        &   25.3 & 1.6 $\times 10^{-6}$ & 1300                \\
    CRESST M2
        &   11.6 & 3.7 $\times 10^{-5}$ & 3.4 $\times 10^{4}$ \\
    DAMA best-fit
        &   11.0 & 2.0 $\times 10^{-4}$ & 1.8 $\times 10^{5}$ \\
    XENON allowed
        &   60.  & 1.  $\times 10^{-8}$ & 5.4 \\
    XENON allowed
        & 1000.  & 8.  $\times 10^{-8}$ & 3.9 \\
    \hline 
    \multicolumn{4}{c}{\textit{spin-dependent}} \\
    DAMA best-fit
        &   11.0 & 0.68                      &  9.0 $\times 10^{4}$ \\
    XENON allowed
        &   60.  & 0.001                     & 30. \\
    XENON allowed
        & 1000.  & 0.01                      & 19. \\
    \hline \hline 
  \end{tabular}
  \end{center}
  \caption[Benchmark points]{%
    The total number of scatters within a human body per year for
    the given WIMP masses and WIMP-proton scattering cross-sections.
    The CoGeNT, CRESST, and DAMA benchmarks are those that best fit
    the data for the respective experiments (CRESST has two maximum
    likelihood points);
    these points are all strongly disfavored by the null results of
    CDMS and XENON in the standard framework used in this analysis.
    The XENON benchmarks are compatible with the null results of
    CDMS and XENON.
    We assume a human mass of 70~kg and identical couplings to the
    proton and neutron.
    }
  \label{tab:Benchmarks}
\end{table}

In \reftab{Benchmarks}, we show scattering rates in the body for
several WIMP benchmarks.  The benchmarks are chosen to correspond to
the approximate best-fit WIMP mass and scattering cross-section for
the CoGeNT \cite{Aalseth:2010vx,Aalseth:2011wp},
CRESST \cite{Angloher:2011uu}, and DAMA \cite{Bernabei:2010mq}
experiments.  Two CRESST benchmark points are included, corresponding
to the two sets of parameters that maximize their likelihood function,
M1 (the global maximum) and M2 (a local maximum).  While DAMA likewise
has two best-fit points, we have included only the lower mass one as
the higher mass point is in strong conflict with the null results of
XENON \cite{Angle:2009xb,Aprile:2011hi} and CDMS \cite{CDMSII}.
We note that, in fact, all of the CoGeNT, CRESSST, and DAMA benchmark
points are incompatible with XENON and CDMS under the analysis framework
we are using here.  Many researchers are trying to understand the origin
of these differences; in this paper we simply follow the published
results in choosing our benchmark points.
Two additional benchmark points are included, corresponding
to the maximum cross-section consistent with the null results of XENON
(and CDMS, which has a slightly weaker constraint) for WIMP masses of
60~GeV and 1~TeV; the former case is the benchmark used in
\reftab{Elements}.
All benchmarks are included for the SI case, while only the DAMA
best-fit and XENON-allowed benchmarks are included in the SD case.

The scattering rates for the CoGeNT, CRESST, and DAMA benchmark points
are all significantly larger than the rates for the XENON-allowed
benchmarks, as the former are all at cross-sections higher than those
that would produce the allowed few events/year observed in XENON.  The
rates for these positive-signal benchmarks vary from $\sim$4 per day
(CRESST M1) to $\sim$20 per hour (DAMA, SI case).  For the XENON-allowed
cases, the rates are several per year in the SI case, but a moderately
larger $\sim$2 per month in the SD case.

At WIMP masses below 60~GeV, XENON begins to lose sensitivity: the rate
above threshold becomes a smaller and smaller portion of the total rate.
For low masses, one can thus choose cross-sections resulting in very
large total rates (in both the human body and XENON detector), that
produce only a few events above threshold and are thus not excluded by
XENON.

WIMP interactions represent a source of radiation in the human body, so
a question arises: are WIMP collisions dangerous to humans?  Here
we compare the radiation due to WIMPs with that from natural  sources,
namely radioactivity here on Earth (including radon) as well as cosmic
rays coming down through the atmosphere.
The natural radiation background  varies by location, with a typical
annual exposure of 0.4--4~mSv (see Refs.~\cite{Nakamura:2010zzi,Pochin:1983}
for a review; here the unit of radiation exposure is Sieverts, or Sv).  
The cosmic-ray contribution is 0.3~mSv/yr at sea level and increases at
higher elevations.
Cosmic-ray muons deposit far more energy in the human body than do WIMPs.
These muons pass through the human body at a rate of a few per
second, depositing $\sim$ 10--100~MeV of energy each, far larger than
the $\sim$ 10~keV deposited by a WIMP.  For comparison, for the
XENON-allowed benchmarks we have considered, the dose-equivalent
exposure due to WIMP interactions is $\orderof{10^{-11}}$~mSv/yr, a
negligible exposure compared to other natural radiation sources.
Indeed we find that the radiation dose from cosmic-rays received
each second exceeds the lifetime WIMP dose.  Even for the higher WIMP
interaction rates for the masses and cross-sections that can
reproduce the CoGeNT, CRESST, and DAMA results, the WIMP radiation dose
is negligible compared to other radiation sources.  Thus WIMPs are
harmless to the human body.


In conclusion, we have studied the interactions of WIMPs with nuclei in
a human body of mass 70~kg.  We examined the contributions from a
variety of elements in the body and found that the dominant
contribution is from scattering off of oxygen nuclei for
spin-independent (SI) interactions and hydrogen nuclei for
spin-dependent (SD) interactions.
For a canonical case of 60~GeV WIMP mass and the maximum elastic
scattering cross-sections compatible with the experimental bounds
from XENON and CDMS ($\sigmapSI = 10^{-8}$~pb = $10^{-44}$~cm$^2$ and
$\sigmapSD = 2 \times 10^{-3}$~pb), we found that on average five WIMPs
hit one of the nuclei in the human body in a year via SI scattering and
30 via SD scattering. 
We also studied the 10--20 GeV WIMPs with much larger cross-sections
that best fit the DAMA, COGENT, and CRESST data sets, and found much
higher rates: in this case as many as $10^5$ WIMPs hit a nucleus in the
human body in an average year, corresponding to almost one a minute.
Finally, we have determined that, while these WIMP interactions
represent a source of radiation in the body, the exposure rate is
negligible compared to that from other natural sources of radiation
and WIMP collisions are harmless to humans.


\begin{acknowledgments}
  K.F.\ acknowledges the support of the DOE and the Michigan
  Center for Theoretical Physics via the University of Michigan.
  K.F.\ thanks the Caltech Physics Dept for hospitality during her visit.
  C.S.\ is grateful for financial support from the Swedish Research
  Council (VR) through the Oskar Klein Centre.
\end{acknowledgments}




\end{document}